\documentclass[10pt]{article}

\usepackage[utf8]{inputenc} 
\usepackage[T1]{fontenc}
\usepackage[english]{babel}
\usepackage{csquotes}
\usepackage[cmex10]{amsmath}  
\usepackage{amsfonts}
\usepackage{amssymb}
\usepackage[a4paper]{geometry}

\newtheorem{theorem}{\textbf{Theorem}}
\newtheorem{proposition}[theorem]{\textbf{Proposition}}
\newtheorem{lemma}[theorem]{\textbf{Lemma}}

\newcommand{\Z}{\mathbb{Z}}
\newcommand{\F}{\mathbb{F}}
\newcommand{\bC}{\mathbf{C}}
\newcommand{\bH}{\mathbf{H}}
\newcommand{\bM}{\mathbf{M}}
\newcommand{\bm}{\mathbf{m}}
\newcommand{\bs}{\mathbf{s}}

\newcommand{\bu}{\mathbf{u}}
\newcommand{\bv}{\mathbf{v}}

\newcommand{\bx}{\mathbf{x}}
\newcommand{\by}{\mathbf{y}}
\newcommand{\wt}{\mathsf{wt}}
\newcommand{\qed}{\phantom{.} \hfill $\square$}

\begin{document}

\title{Upper Bounds on the Minimum Distance of Structured LDPC Codes}

\author{François Arnault\thanks{XLIM, UMR 7252, Université de Limoges, France \\ Emails: \textbf{\{arnault, philippe.gaborit, nicolas.saussay\}@unilim.fr}} \and Philippe Gaborit\footnotemark[1] \and Wouter Rozendaal\thanks{IMB, UMR 5251, Université de Bordeaux, France \\ Emails: \textbf{\{wouter.rozendaal, gilles.zemor\}@math.u-bordeaux.fr}} \and Nicolas Saussay\footnotemark[1] \and Gilles Zémor\footnotemark[2] \thanks{Institut Universitaire de France}} 

\maketitle


\begin{abstract}
We investigate the minimum distance of structured binary Low-Density Parity-Check (LDPC) codes whose parity-check matrices are of the form $[\bC \, \vert \, \bM]$ where $\bC$ is circulant and of column weight $2$, and $\bM$ has fixed column weight \mbox{$r \geq 3$} and row weight at least $1$. These codes are of interest because they are LDPC codes which come with a natural linear-time encoding algorithm. We show that the minimum distance of these codes is in $O(n^{\frac{r-2}{r-1} + \epsilon})$, where $n$ is the code length and $\epsilon > 0$ is arbitrarily small. This improves the previously known upper bound in $O(n^{\frac{r-1}{r}})$ on the minimum distance of such codes. 
\end{abstract}


\section{Introduction}

In this short note, we are interested in binary LDPC codes whose parity-check matrices are of the form $[\bC \, \vert \, \bM]$ where $\bC$ is circulant with exactly two non-zero elements per column and $\bM$ has constant column weight $r \geq 3$. Such codes were introduced in \cite{TZ06}, where the point was made that the matrix $\bC$ ensured a number of variable nodes of degree $2$ (i.e. incident to two parity-check nodes) equal to the total number of parity-check equations which was known to be a feature of capacity-achieving LDPC codes \cite{YRL04, HEA05, R03}. More importantly though, the main motivation for introducing the matrix $\bC$ with a circulant structure is that it ensures a simple and natural linear time encoding algorithm \cite{M03, YRL04, TZ06} that vanilla LDPC codes lack. The interest for linear-time encodable LDPC codes was rejuvenated lately by multiparty computation applications \cite{B22} where the LDPC property is useful for speedy computations. The minimum distance of such codes is relevant to security considerations and it is the primary focus of the present paper.

The minimum distance of these structured LDPC codes was studied in \cite{TZ06}. It was shown that when the matrix $\bM$ has constant row weight, and constant column weight $r$, then the typical minimum distance of the resulting code scales as $\Omega(n^{(r-2)/r})$, where $n$ is the blocklength. An upper bound on the minimum distance was also proved which was of the form $O(n^{(r-1)/r})$. This upper bound makes no real use of the row constraints so is valid for any matrix $\bM$ that has constant column weight $r$ and has a number of columns at least equal to a constant fraction of that of~$\bC$. We improve this upper bound and show that their minimum distance is in $O(n^{\frac{r-2}{r-1} + \epsilon})$, with $\epsilon > 0$ arbitrarily small, thus narrowing the gap between upper and lower bounds. The upper bound of \cite{TZ06} embeds the columns of $\bM$ into a metric space and uses a packing argument: here we embed sums of columns of $\bM$ into a similar metric space, but the sums of columns need to be carefully chosen, namely in a {\em quasi-colliding} manner that we introduce below.


\section{Structured LDPC codes: Main theorem}

The family of binary LDPC codes we consider are the codes with $m \times n$ parity-check matrices of the form $\bH = [\bC \, \vert \, \bM]$ where $\bC$ is the $m \times m$ circulant matrix
\[\bC = \begin{pmatrix}
1 &   &        &   & 1 \\
1 & 1 &        &   &   \\
  & 1 & \ddots &   &   \\
  &   & \ddots & 1 &   \\
  &   &        & 1 & 1
\end{pmatrix}\]
and $\bM$ is a binary $m \times (n-m)$ matrix such that it has no $\mathbf{0}$ row and all of its columns have weight $r \geq 3$. One readily checks that such codes exist if and only if 
\begin{equation}
\label{eq:existence condition}    
m \geq r \text{ and } (n-m)r \geq m. 
\end{equation}

Equivalently, the family of $(n,m,r)$-structured LDPC codes we consider is the ensemble of codes with Tanner graph chosen to have $m$ check nodes of weight $\geq 3$, $m$ variable nodes of degree $2$ that lie on a single cycle, and all other $n-m$ variable nodes of constant degree $r \geq 3$. 

We are interested in the asymptotic behaviour of the minimum distance of such codes when the column parameter $r$ is fixed and the blocklength $n$ tends to infinity. Our main result states:

\begin{theorem}
\label{thm:classical upper bound}
For a fixed column weight parameter $r \geq 3$, and large blocklength $n$, the minimum distance $d_{\min}$ of any binary $(n,m,r)$-structured LDPC code is smaller than a quantity of order $O(n^{\frac{r-2}{r-1}+\epsilon})$, where $\epsilon > 0$ is fixed and arbitrarily small.
\end{theorem}

\section{Proof strategy}

Suppose that $C$ is a $(n,m,r)$-structured LDPC code defined by a parity matrix of the form $\bH = [\bC \, \vert \, \bM]$. Notice that the sum of $t$ consecutive columns of $\bC$ yields a column vector of weight $2$ and support $\{i,j\}$ with $\ell(i-j)=t$, where $\ell(x)$ denotes the smallest absolute value of an integer $y$ such that $x = y$ mod $m$. This suggests the following definition: for a vector $\bv$ with support $s$, a pair of coordinates $\{i,j\}$ of $s$ will be said to be a \textit{quasi-colliding} pair of tolerance $t$ for $\bv$, or for $s$, whenever $\ell(i-j)\le t$. We have:

\begin{proposition}
\label{prop:idea of proof}
If there exists a set $S$ of columns of the matrix $\bM$ that sums to a column vector whose support can be partitioned into pairs of quasi-colliding coordinates of tolerance $t$, then \mbox{$d_{min} \leq |S| + t|S|r/2$}. 
\end{proposition}
\noindent
\textit{Proof:} Since every column of $\bM$ has weight $r$, the sum $\bs$ of the columns in $S$ has Hamming weight at most $|S|r$ and decomposes into at most $|S|r/2$ quasi-colliding pairs. Therefore, $\bs$ can also be written as at most $|S|r/2$ sums of at most $t$ consecutive columns of $\bC$. \qed \\

The general strategy for deriving an upper bound on the minimum distance is therefore to prove the existence of a small set $S$ of columns for which Proposition~\ref{prop:idea of proof} applies with the smallest possible tolerance parameter $t$.

In \cite{TZ06}, the authors bounded the minimum distance of such LDPC codes by taking $|S| = 2$. In the present paper, $|S|$ is constant and arbitrarily large, and $t$ is growing sublinearly in $n$. $S$ is obtained from two sets $S_1$ and $S_2$ of columns of $\bM$. The sums of the columns in each of those sets, $\mathbf{s}_1$ and $\mathbf{s}_2$ respectively, will be such that they contain a predefined number of built-in quasi-colliding coordinates. The other quasi-collisions will then be obtained by pairing the remaining coordinates of $\mathbf{s}_1$ and $\mathbf{s}_2$ together. Such a pairing will follow from a packing argument, using the fact that there are sufficiently many choices for $\mathbf{s}_1$ and $\mathbf{s}_2$.

\section{Proof of Theorem~\ref{thm:classical upper bound}}

\subsection*{Notation and definitions} 

Notice that the columns of $S$ sum to a column vector of Hamming weight typically equal to $|S|r$. It may be less however, if there are collisions between the supports of columns of $S$. To avoid dealing with these special cases, we will sum vectors of $\bM$ over $\Z$ rather than $\F_2$. Consequently, these vectors will have multisupports rather than supports, where coordinates appear multiple times. For $\bu, \bv \in \Z_{\geq 0}^{m}$, not necessarily distinct, we say that there is a quasi-collision between $\bu$ and $\bv$ of tolerance $t \in \Z_{\geq 0}$ if there exists $(i,j) \in s_{\bu} \times s_{\bv}$ such that $\ell(i-j) \leq t$. As noted above, the pair of quasi-colliding coordinates can be expressed by less than $t$ columns of the circulant matrix $\bC$. 

For $\bv \in \Z_{\geq 0}^{m}$, we define the weight of $\bv$ to be the non-negative integer \mbox{$\wt(\bv) = \sum_{i=1}^{m} v_{i}$}. Let \mbox{$V_{c} = \{ \bv \in \Z_{\geq 0}^{m} \vert \sum_{i=1}^{m} v_{i} = c \}$} be the set of $m$-tuples of non-negative integers of weight $c \in \Z_{\geq 0}$. For $\bu, \bv \in V_{c}$, we define a pairing between $\bu$ and $\bv$ to be a set $P_{\bu,\bv}$ of $c$ ordered pairs \mbox{$(i,f(i)) \in s_{\bu} \times s_{\bv}$}, where $f$ is a bijection between the multisupports $s_{\bu}$ and $s_{\bv}$. Given a pairing $P_{\bu,\bv}$ between $\bu$ and $\bv$, we define the \textit{width} of the pairing to be the non-negative integer \mbox{$|P_{\bu,\bv}| = \max \{ \ell(i-j) \vert (i,j) \in P_{\bu,\bv} \}$.} 

Let us denote by $M_k \subset \Z_{\geq 0}^{m}$ the set of column vectors obtained by summing $k+1$ distinct columns of the matrix $\bM$ and having at least $k$ built-in quasi-colliding pairs. The weight of $\bv \in M_k$ is $(k+1)r$. By disregarding $k$ built-in quasi-colliding pairs, one can consider the $m$-tuple $\tilde{\bv} = \bv - \bx$, where $\bx$ is a sum of not more than $tk$ columns of $\bC$. The weight of $\tilde{\bv}$ is then equal to $(k+1)r - 2k = k(r-2) + r$, i.e. $\tilde{\bv} \in V_{c}$ for $c = k(r-2) + r$. For every $\bv \in M_k$, we consider the associated reduced $m$-tuple $\tilde{\bv}$ and define the set $B_{\tilde{\bv},t}$ of vectors $\bu \in V_{c}$, with $c = r+(r-2)k$, such that there exists a pairing $P_{\tilde{\bv},\bu}$ between $\tilde{\bv}$ and $\bu$ of width less than $t/2$. \\

We can now spell out the core of the argument and show how to make use of these quasi-collisions to derive an upper bound on the minimum distance of the code.

\begin{lemma}
\label{lem:metric argument}
If there exist two vectors $\bu, \bv \in M_k$ such that the sets $B_{\tilde{\bu},t}$ and $B_{\tilde{\bv},t}$ intersect non-trivially, then there exists a non-zero codeword of Hamming weight less than or equal to \mbox{$2(k+1) + t(k+1)r$}.
\end{lemma}
\noindent
\textit{Proof:} The vectors $\tilde{\bu}, \tilde{\bv}$ are by construction equal to $\bu - \bx$ and $\bv - \by$ respectively, where $\bu, \bv$ are sums of $k+1$ columns of $\bM$, and $\bx,\by$ are sums of less than $tk$ columns of $\bC$. If $B_{\tilde{\bu},t}$ and $B_{\tilde{\bv},t}$ intersect nontrivially, then there exists a pairing $P_{\tilde{\bu},\tilde{\bv}}$ of width less than $t$ between $\tilde{\bu}$ and $\tilde{\bv}$. Hence $\tilde{\bu} + \tilde{\bv}$ can be written as a sum of at most $tc$ columns of $\bC$, with $c = k(r-2) + r$. We thus have $\bu + \bv - (\tilde{\bu} + \tilde{\bv} + \bx + \by) = 0$, with $(\bu + \bv)$ a sum of at most $2(k+1)$ columns of $\bM$ and $(\tilde{\bu} + \tilde{\bv} + \bx + \by)$ a sum of at most $tc + 2tk$ columns of $\bC$. It follows that there is a non-zero codeword whose Hamming weight is at most $2(k+1) + t(c+2k) = 2(k+1) + t(k+1)r$. \qed \\

By Lemma~\ref{lem:metric argument}, it suffices to find two vectors $\bu, \bv \in M_k$ such that $B_{\tilde{\bu},t}$ and $B_{\tilde{\bv},t}$ intersect non-trivially in order to obtain an upper bound on the minimum distance of the code. The existence of two such vectors is guaranteed if 
\begin{equation}
\label{eq:packing condition}
\sum_{\bv \in M_k} \vert B_{\tilde{\bv},t} \vert > \vert V_{c} \vert.
\end{equation}

We therefore need to evaluate the quantities $\vert M_k \vert$, $\vert B_{\tilde{\bv},t} \vert$ and $\vert V_{c} \vert$. We start by providing the following upper bound on the size of $V_{c}$, the set of $m$-tuples of non-negative integers of weight $c$.

\begin{lemma}
\label{lem:V_c}
$\vert V_{c} \vert \leq \frac{(m+c)^c}{c!}$.
\end{lemma}
\noindent
\textit{Proof:} There are exactly as many elements in $V_c$ as ways to choose $c$ elements from a set of $m$ elements when repetitions are allowed, which equals $\tbinom{m+c-1}{c}$, from which we straightforwardly get $\vert V_{c} \vert \leq \frac{(m+c)^c}{c!}$. \qed \\

We now give, for any given $\bv \in M_k$, the following lower bound on the size of $B_{\tilde{\bv},t}$, the set of vectors $\bu \in V_{c}$ such that there exists a pairing $P_{\tilde{\bv},\bu}$ between $\tilde{\bv}$ and $\bu$ of width less than $t/2$.

\begin{lemma}
\label{lem:B_v,t}
$\vert B_{\tilde{\bv},t} \vert \geq \frac{t^c}{c!}$.
\end{lemma}
\noindent
\textit{Proof:} Let us count how many vectors $\bu \in B_{\tilde{\bv},t}$ we can construct. A pairing between $\tilde{\bv}$ and $\bu$ corresponds to finding a bijection between the multisupports $s_{\tilde{\bv}}$ and $s_{\bu}$. If we want a pairing of width less than $t/2$, then we can construct $\bu$ by choosing, for each $i \in [[1,m]]$, $\tilde{v}_{i}$ elements among the $t+1$ possible positions around the index $i$. If these sets of possible positions overlap, then we may construct the same vector in different ways. In the worst case $s_{\tilde{\bv}}$ consists of only one non-zero entry, and thus there are always at least as many elements in $B_{\tilde{\bv},t}$ as ways to choose $c$ elements among $t+1$ elements when repetitions are allowed. Therefore $\vert B_{\tilde{\bv},t} \vert \geq \tbinom{t+c}{c} \geq \frac{t^c}{c!}$. \qed \\

Let us now provide the following lower bound on the size of $M_k$, the set of $m$-tuples obtained by summing $k+1$ distinct columns of $\bM$ and having at least $k$ built-in quasi-colliding pairs.

\begin{lemma}
\label{lem:M_k}
If $t > 2k$, then $\vert M_k \vert > \frac{n-m}{(k+1)!}t^{k}$.
\end{lemma}
\noindent
\textit{Proof:} To bound the size of $M_k$ we will construct $m$-tuples of $M_k$ by considering $k$ successive quasi-collisions of distinct columns of the matrix $\bM$. More precisely, we start by choosing a column $\bm_{i_0}$ of $\bM$. The first column may be chosen among all the columns of $\bM$ so there are $n-m$ choices for $\bm_{i_0}$. We then pick another column $\bm_{i_1}$ of $\bM$ such that there is a quasi-collision between $\bm_{i_1}$ and the first column. To this end, we choose one of the non-zero entries of $\bm_{i_0}$, say the $j$-th entry. Next, to have a quasi-collision of tolerance $t$, we choose one of the $2t$ rows of $\bM$ whose index $i \neq j$ satisfies $\ell(i-j) \leq t$. Once a row is chosen, we pick one of the columns whose entry in that row is non-zero. Recall that all rows of $\bM$ are non-zero. If this column is distinct from the first column, then we take it to define $\bm_{i_1}$. Otherwise, if the chosen column is equal to the first column, then there is already a quasi-collision between the coordinates of $\bm_{i_0}$. We may thus chose $\bm_{i_1}$ to be any other column. It follows that there are at least $2t$ valid choices for $\bm_{i_1}$. We then pick another column $\bm_{i_2}$ of $\bM$ such that there is a quasi-collision between $\bm_{i_2}$ and the sum $\bm_{i_0} + \bm_{i_1}$. We now choose one of the non-zero entries of this sum whose support isn't used for a quasi-collision, and consider one of $2t-2$ rows of $\bM$ whose index can give rise to another quasi-collision. Once a row is chosen, we pick one of the columns whose entry in that row is non-zero. If this column is distinct from the two first columns, then we take it to define $\bm_{i_2}$. Otherwise, there is a quasi-collision between the coordinates of $\bm_{i_0}+\bm_{i_1}$. We may thus choose any other column to define $\bm_{i_2}$. It follows that there are at least $2t-2$ valid choices for $\bm_{i_2}$. Repeating this argument, it is clear that $\forall \, q \in [[1,k]]$, there are at least $2t - 2(q-1)$ choices for $\bm_{i_q}$. We therefore have \[\vert M_k \vert \geq \frac{n-m}{(k+1)!} \prod_{i=0}^{k-1}(2t-2i) \geq \frac{n-m}{(k+1)!}(2t-2k)^{k}.\] 
Whenever $t > 2k$, we have $2t -2k > t$ and so we obtain $\vert M_k \vert > \frac{n-m}{(k+1)!}t^{k}$. \qed \\

Putting everything together, we now proceed to prove the main theorem. \\

\noindent
\textit{Proof of Theorem \ref{thm:classical upper bound}:} 
Let $r \geq 3$ and $k \in \Z_{\geq 0}$ be fixed, let $c = (r-2)k + r$ and $t > 2k$. Consider a $(n,m,r)$-structured LDPC code $C$. From condition \eqref{eq:packing condition} and the inequalities in Lemma~\ref{lem:V_c}, \ref{lem:B_v,t} and \ref{lem:M_k}, the existence of a non-zero codeword of Hamming weight less than or equal to \mbox{$t(k+1)r+2(k+1)$} is guaranteed if
\begin{equation} 
\label{eq:packing condition 2}
(n-m) t^{k+c} \frac{1}{(k+1)!} \geq (m+c)^{c}. 
\end{equation}
From condition~\eqref{eq:existence condition}, the inequality \eqref{eq:packing condition 2} is satisfied if 
\begin{equation} 
\label{eq:packing condition 3}
t^{k+c} \geq r(k+1)! \frac{(m+c)^{c}}{m}.
\end{equation}
We now choose $t$ as small as possible whilst respecting condition \eqref{eq:packing condition 3}, namely \[t = \left\lceil\left( r(k+1)! \frac{(m+c)^{c}}{m} \right)^{1/k+c}\right\rceil.\] 
Lemma~\ref{lem:metric argument} now gives us a non-zero codeword of weight less than or equal to \mbox{$t(k+1)r+2(k+1)$}. Hence, remembering that $r$, $k$ and $c$ are constants, we get, for growing $m$, \[ d_{\min} = O(m^{\frac{c-1}{k+c}}) = O(m^{\frac{r-2}{r-1} + \epsilon})\] with $\epsilon = 1/(r-1)(k(r-1)+r)$. The constant $\epsilon$ can be made arbitrarily small by choosing $k$ large enough. Finally, since $m \leq n$, we conclude that asymptotically, $d_{\min} = O(n^{\frac{r-2}{r-1} + \epsilon})$.
\qed


\section*{Acknowledgements}
All authors acknowledge the support from the Plan France 2030 through the project NISQ2LSQ, ANR-22-PETQ-0006. 


\newcommand{\etalchar}[1]{$^{#1}$}

\end{document}